\documentclass[aps,preprint,onecolumn]{revtex4}
\usepackage{graphicx}
\usepackage{dcolumn}
\usepackage{bm}

\begin{document}
\preprint{PREPRINT}
\title{How to erase surface plasmon fringes}
\author{Aurelien Drezet}
\email{aurelien.drezet@uni-graz.at}
\author{Andrey L. Stepanov, Andreas~Hohenau, Harald Ditlbacher, Bernhard
Steinberger, Nicole Galler, Franz R. Aussenegg, Alfred Leitner,
Joachim R. Krenn} \affiliation{Institute of Physics and Erwin
Schr\"{o}dinger Institute for Nanoscale Research,
Karl-Franzens-University Graz, Universit\"{a}tsplatz 5, A-8010
Graz, Austria}

\begin{abstract}
We report the realization of a dual surface plasmon polariton
(SPP) microscope based on leakage radiation (LR) analysis. The
microscope can either image SPP propagation in the direct space or
tin the Fourier space. This particularity allows in turn
manipulation of the LR image for a clear separation of different
interfering SPP contributions present close to optical
nanoelements .
\end{abstract}



\maketitle

The miniaturization of optical elements and devises into nanoscale
dimensions is restricted by the diffraction limit to about half of
the effective light wavelength. One promising way to avoid this
restriction is the use of surface plasmon polaritons (SPPs)
instead of light waves. SPPs are quasi-two-dimensional
electromagnetic waves of electron excitations, propagating at a
metal-dielectric interface and having field components decaying
exponentially into both neighboring media \cite{Raether}. As was
recently demonstrated \cite{Hecht,Bouhelier,Stepanov}, to image
the spatial SPP profile, besides near-field optical microscopy
\cite{Weeber} or fluorescence mapping \cite{Harry}, leakage
radiation (LR) imaging microscopy can be applied. It was shown
that this new approach allows for quantitative measurements of the
spatial SPP field profile by deducing SPP reflection,
transmission, and scattering efficiencies for various surface
nanostructures \cite{Stepanov,Drezet}. In parallel to this LR
microscopy in the direct space it has been recently experimentally
demonstrated that LR imaging is equivalently possible in the
Fourier space~\cite{Barnes},e.~g., imaging in the SPP wavevector
space. Here, based on the use of a dual LR microscopy working in
both the direct and Fourier space, we present the next development
of LR imaging microscope and we discuss new possibilities for
imaging and controlling of SPPs. In particular we show how by
acting in the Fourier space this method can be applied to erase
from the final image in the direct space SPP interferences fringes
existing close to structure like Bragg mirror \cite{Harry,Drezet}.
This in turn allows quantitative analysis in a spatial region
where near field optics can not resolve and distinguish the
different SPP field components contributing to the SPP image.
\begin{figure}[h]
\includegraphics[width=3in]{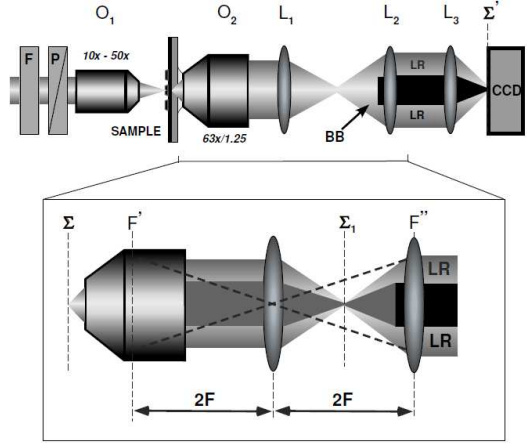}\\
\caption{Experimental scheme for Leakage Radiation (LR) imaging.
SPPs are exited by laser light on a structured gold film on a
glass substrate. LR is emitted into the glass substrate at an
angle $\theta_{LR}$. F, gray filter; P, polarizer; BB, beam
block.}
\end{figure}

The intensity decay length of a plane SPP wave in a perfectly
planar metal film between two dielectric media defines its
intrinsic decay length $L_{\textrm{SPP}}=1/2k"_{\textrm{SPP}}$,
which is a measure of the "ideality" of the electron gas.
$k"_{\textrm{SPP}}$ is defined as the imaginary part of the
complex SPP wave vector
$k_{\textrm{SPP}}=k'_{\textrm{SPP}}+ik"_{\textrm{SPP}}$. Intrinsic
losses are caused by inelastic scattering of conduction electrons,
scattering of electrons at interfaces and leakage radiation LR
\cite{Raether}. LR is emitted from the interface between a metal
thin film and the higher refractive index medium (glass substrate)
\cite{Stepanov,Drezet}. When the electromagnetic SPPs field cross
the metal film and reach the substrate, LR appears at a
characteristic angle of inclination $\theta _{LR}$ with respect to
the interface normal. At this angle the LR wave satisfies
$k'_{\textrm{SPP}}=2\pi/\lambda_{SPP}=n k_{0}$sin$\theta_{LR}$,
where $nk_{0}$ is the wave vectors of LR, $n$ being the refractive
index of the glass substrate, and $\lambda_{SPP}$ the SPP
wavelength. Although LR contributes to SPP damping, it permits the
direct mapping of the SPP propagation at the air/metal interface.
Indeed, the intensity collected at any point $P'$ of the image
plane $\Sigma'$ of the LR microscope with a charged-coupled-device
(CCD) camera is directly proportional to the intensity of SPPs at
the conjugate point $P$ located on the air/metal boundary, i.~e.~,
in the object plane $\Sigma$ of the microscope.
\cite{Stepanov,Drezet}.
\begin{figure}[h]
\includegraphics[width=3in]{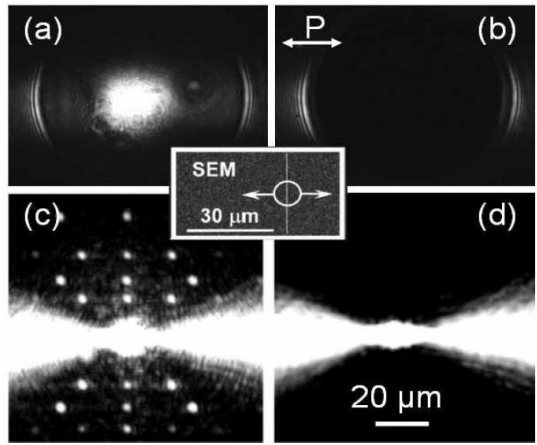}\\
\caption{LR Fourier plain images measured with blocked central
beam (left) and without blocking (right). P shows the laser
polarization direction.}
\end{figure}
The new LR microscope with the improvements discussed in the
following is sketched on Fig.~1.  Like in the previous version of
the LR microscope \cite{Stepanov,Drezet} and in order to avoid
total internal reflection inside the glass substrate, an oil
immersion objective $O_2$ 63$\times$, numerical aperture of 1.25)
in contact with the bottom part of the sample is used to collect
LR images.  It is important to remark that simultaneously to LR
the incident exiting laser beam is going through the sample and
the microscope contributing subsequently to the total signal in
the $\Sigma'$ plane. This direct laser light dominates the SPP
signal in the region of excitation in $\Sigma$. This feature is
conserved in $\Sigma'$ where some distortions due to saturation of
the camera are even added to the resulting image. It is
consequently impossible to separate SPPs propagation from the
incident beam in the vicinity of the excitation region. However,
as well known from Fourier optics one can distinguish these two
contributions by observing the intensity distribution in the back
focal plane $F'$ of the immersion objective. Indeed since the
signal collected in $F'$ is a cartography of the 2D momentum
distribution of photons emitted in $\Sigma$ the SPP distribution
is thus confined on a circle corresponding to $k'_{SPP}$.
Additionally, the central beam (with an angular divergence much
smaller than $\theta_{LR}$) is located at the center of this
circle. By introducing a central beam-block in $F'$ it is then
possible to eliminate the direct contribution of the laser beam.
By contrast with the previous version of LR microscope
\cite{Stepanov} we included this beam block in the optical setup.
Since however $F'$ is contained in the immersion objective we must
first image $F'$, in a plane
 $F''$, by using a lens $L_{1}$ in a 2f-2f configuration. In $F''$ we can introduce the central beam-block (see Fig.~1). In
order to image the SPP propagation in the direct space we
introduce two auxiliary lenses $L_2$ and $L_3$ focusing light on
the CCD ($L_2$ is located in $F''$). By changing the focal length
of $L_3$ one can either image the Fourier plane $F''$ or the
object plane $\Sigma_1$ (image of $\Sigma$ through $O_2$ and
$L_1$). This dual microscope is consequently able of recording
SPPs propagation either in the direct or in the Fourier space.

\begin{figure}[h]
\includegraphics[width=3in]{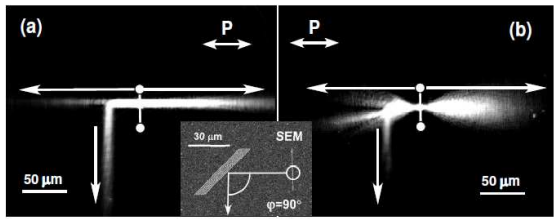}\\
\caption{LR images of SPPs launched from a ridge and interacting
with a Bragg mirror (SEM structure in the inset). SPPs impinge at
$45^{\circ} $incidence on the Bragg mirror (optimized for this
angle at $\lambda=800$ nm). (a) Image with $O_1$= (10$\times$,
numerical aperture 0.25). (b) Image with $O_1$= (50$\times$,
numerical aperture 0.7). P shows the laser polarization
direction.}
\end{figure}
In order to illustrate the potentiality of this optical setup we
consider different nanostructures obtained by usual electron beam
lithography (EBL) \cite{Krenn} on a 60 nm thick gold film. For the
local excitation of SPPs we focus a linearly polarized light from
a Ti:sapphire laser  (wavelength: $\lambda=800\simeq\lambda_{SPP}$
nm) through a microscope objective $O_1$ onto a gold ridge (200 nm
wide, 60 nm high). Launched SPPs propagate to the left and to
right of the ridge in the direction perpendicular to the ridge
axis. Figs.~2a, b, c, d show LR images of SPPs corresponding to
this configuration ($O_1$:50$\times$, numerical aperture 0.7).
Figs.~2a and c correspond to imaging in the Fourier and direct
space respectively. The central spot associated with the laser
beam and the partial ring associated with the SPPs cone are
clearly visible on Fig.~2a. This laser beam saturates the image in
the direct space and creates some artefact in the launching region
as visible on Fig.~2c. Figs.~2b,d show the same images with the
central beam block introduced in the plane $F'$. As a consequence
of this introduction the multiple artefact disappear from the
image in the direct space resulting in improvements of image
quality and eliminating the spurious effect of the incident laser
light (see Fig.~2d).

As a further modification, we introduced the use of different
objectives for focusing of the laser beam. The different focus
diameters achievable with the various objectives of focal lenght
$f_{1}$ allow in turn the generation of SPP waves of different
divergence angles. Indeed the convergence angle $\alpha$ of the
laser beam on the sample is connected to the laser beam diameter
$W\simeq 2$ mm just before the lens by the relation
$\tan{(\alpha)}=W/(2f_{1})$. Due to Heisenberg's relation we have
in the sample plane $\tan{(\alpha)}=2\lambda/(\pi W_0)$ where
$W_0$ is the focus beam diameter in the sample plane. Since SPPs
launched from the ridge obey to the same Heisenberg relation we
conclude that the divergence angle of the SPP beam equals
$\arctan{[(\lambda/\lambda_{SPP})\cdot\tan{(\alpha)}]}\simeq
\alpha$ , i.~e.~the convergence angle of the laser beam. This
principle is illustrated in Fig.~3 which shows LR images of SPPs
launched by a ridge and reflected by a Bragg mirror \cite{Harry,
Drezet,Weeber}. The Bragg mirror considered here has been
optimized for $45^{\circ}$ incidence angle with respect to the
direction normal to Bragg's mirror in the sample plane. The mirror
is made of gold ridges (60 nm height, 140 nm wide) separated by a
distance $\lambda_{SPP}/\sqrt{2}\simeq 556 nm$ as described in
\cite{Weeber}. Comparison of Fig.~3a ($O_1$: 10$\times$, NA=0.25)
with Fig.~3b ($O_1$: 50$\times$, NA=0.7) show clearly that the
Bragg mirror is much more efficient with a parallel beam
($\alpha=2^{\circ}$, see Fig.~3a) that with a divergent beam
($\alpha=18^{\circ}$, see Fig.~3b). This method of generating a
parallel SPPs beam can be considered as an alternative to prism
technics used in PSTM imaging \cite{Weeber}. The black shadow in
the transmitted beam of Fig.~3b shows directly in counterpart the
reflectivity acceptance angle of the Bragg mirror. Both approach
can be thus useful for understanding SPPs reflectivity of such
system. Additionally it must be added that in order to observe
SPPs reflectivity with a narrow beam like in Fig.~3a the use of
the central beam block is necessary. Indeed without this beam
block the laser beam of diameter $W_0\simeq 15$ $\mu$m would
saturate the recorded signal in the region of interest between the
ridge and the Bragg mirror. The effect is less pronounced for
divergent beam $\alpha=18^{\circ}$ where $W_0\simeq\lesssim 2$
$\mu$m.\\
\begin{figure}[h]
\includegraphics[width=2in]{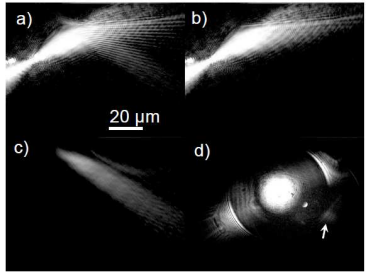}\\
\caption{LR images of SPPs launched from a ridge and interacting
with a Bragg mirror like the one in Fig.~3. SPPs impinge at
$60^{\circ} $incidence on the Bragg mirror ($\lambda=800$ nm). SPP
interference fringes are visible close to the Bragg mirror. (b) LR
image of SPPs propagation in the direct space ($O_1$: 10$\times$,
NA=0.25). (b) LR image of the incident SPP wave launched from the
ridge region. (c) LR  image of the reflected SPP beam. Fringes are
removed from (b) and (c). LR image of SPPs in the Fourier plane
$F''$ without beam block. The introduction of beam blocks in $F''$
(not shown) give rises to figures (b) and (c). The arrow shows the
reflected beam in $F''$.}
\end{figure}
It should be however remarked that the potentiality of such a dual
LR microscope are not limited to correction of artefact or
improvement in LR images quality. Indeed the definition of Fourier
optics itself allow us to manipulate the images in order to
extract some relevant physical information hidden in the pictures.
This is illustrated in Fig.~4 which shows LR images of SPP
reflected at large incidence angle (e.~g.~,
$\theta_{\textrm{inc}}=66^{\circ}$) on a Bragg mirror like the one
shown in Fig.~3. Interference between the incident and reflected
SPP field give rises to fringes in the region close to the mirror
(see Fig.~4a). The presence of interference prohibit a simple and
direct analysis of SPP reflection. However as shown in Fig.~4d the
LR image in the Fourier space separates clearly the respective
contributions of the incoming laser beam (the central spot), of
the incident SPP launched from the ridge (the two arcs of circle),
and of the reflected SPP beam (indicated by an arrow). By
positioning adequately a beam block in this Fourier plane one can
remove not only the incoming laser beam but selectively image
either the incident and transmitted beams (see Fig.~4b) or the
reflected beam (see Fig.~4c) alone. The interference fringes are
clearly erased from the LR pictures. This method in turn allow
direct quantitative analysis not directly possible with the
previous existing methods \cite{Harry,Stepanov,Weeber}.\\
In summary, Thus, based on conventional microscopy dual LR imaging
proves to be a quick and reliable technique for probing SPP fields
with the advantage of providing possible quantitative analysis in
both Fourier and direct space. This dual method is particulary
adapted to analysis of SPP propagation in region where different
beams interfere and where different contributions can thus be
selectively erased for subsequent analysis.\\
The authors thank M.U. Gonz\'{a}lez for test-sample preparation.
For financial support, the European Union under projects FP6
NMP4-CT-2003-505699, FP6 2002-IST-1-507879 and the Lisa Meitner
programm of the Austrian Scientific Foundation (M868-N08) are
acknowledged.

\end{document}